\documentclass{moriond}

\def\be{\begin{equation}}
\def\ee{\end{equation}}
\def\bea{\begin{eqnarray}}
\def\eea{\end{eqnarray}}

\newcommand{\Photo}{\includegraphics[height=35mm]{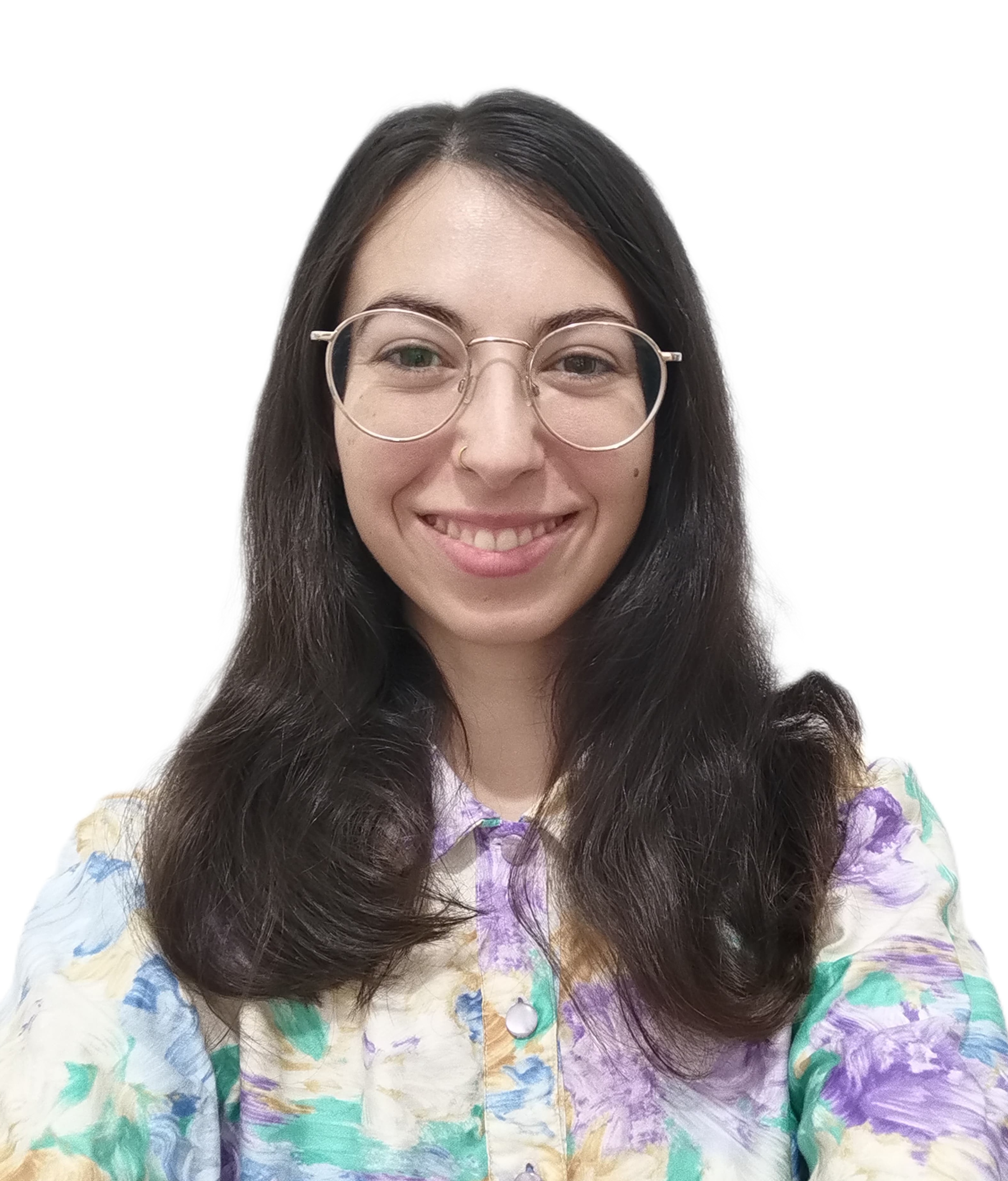}}

\usepackage{graphicx}
\usepackage{subcaption}
\usepackage{caption}

\begin{document}
\vspace*{4cm}
\title{Status of the O4 run and latest non-CBC results}

\author{Martina Di Cesare}

\address{Università di Napoli “Federico II”, I-80126 Napoli, Italy}

\maketitle\abstracts{
The fourth observing run (O4) of Advanced LIGO, Virgo, and KAGRA has started in May 2023 and is planned to continue until October 2025. On behalf of the LVK Collaboration, I will cover two topics: \emph{Status of O4 run} and \emph{latest non-CBC results}.\\
\emph{Status of O4 run}. The focus will be on detectors' performance and online searches/alerts, drawing on publicly available sources provided by the collaboration. Additionally, I will give an overview of removing noise techniques, including AI approaches that help gain sensitivity at a small cost.\\
\emph{Latest non-CBC results}. Compact Binary Coalescence (CBC) is just one of the potential GW sources: Continuous Waves, Bursts, and Stochastic are still being hunted down. Here, O4 public results of searches will be presented, or the latest O3 papers will be discussed when the former are not yet available. \\ So far, no GW detections have been associated with these non-CBC sources in any of the searches conducted.}

\section{Status of the O4 run}

\begin{figure}[ht]
\centering
\includegraphics[width=0.7\linewidth]{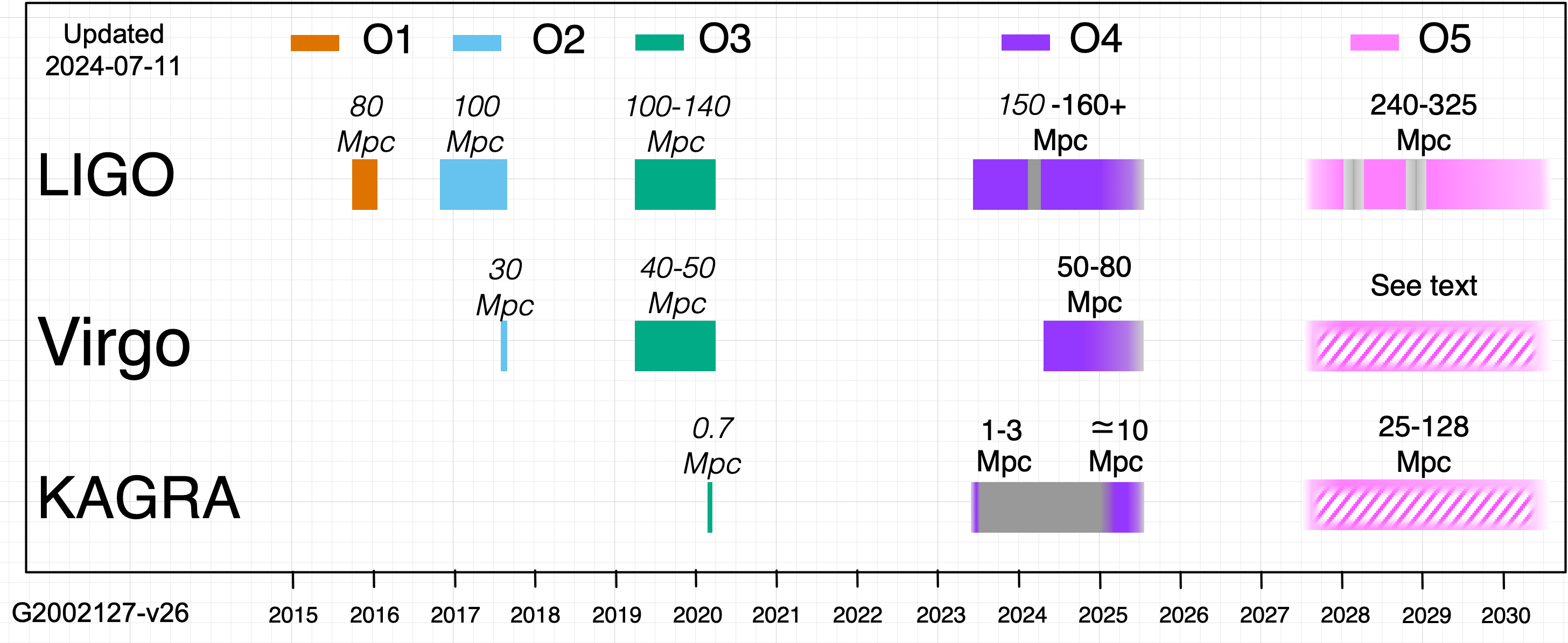}
\caption{Timeline of the Observing Runs \protect\cite{igwn}}
\label{fig:obs_plan}
\end{figure}

The LSC and the Virgo Collaboration have been carrying out joint analyses of available data sets and co-authoring observational result papers since 2010. Beginning in 2021, the KAGRA Collaboration also began co-authoring observational results from the second half of LIGO's third observing run, O3. Together, these three entities comprise the LIGO-Virgo-KAGRA Collaboration (LVK Collaboration). In Fig. \ref{fig:obs_plan}, an overview of the past, present, and future acquisition runs.

\newpage
\subsection{Detectors' performance}
The latest observing run - O4 - has started on 24$^{th}$ May 2023 and is planned to end on 7$^{th}$ October 2025, making it the longest so far. In Table \ref{tab:o4_run}, further details about the sub-intervals (O4a/b/c) are provided. Combining the info of the figure and the table, it's evident the timeline of the three detectors and their contributions: Virgo joined O4 starting from the second run (O4b), due to a \emph{mystery} noise still under investigation; KAGRA was on observing mode for a month during O4a, but then an earthquake occurred in Japan in January 2024; for the LIGO detectors, Livingston had a shutter failure while Hanford experienced - among others - laser glitch issues. Therefore, the gray area describes the period during which each detector was not acquiring data.

\begin{table}[htbp]
    \centering
    \begin{tabular}{|c|c|c|}
    \hline
    Run & Start & End \\
    \hline
         O4a & 2023/05/24  & 2024/01/16  \\
         O4b & 2024/04/10 & 2025/01/28 \\
         O4c & 2025/01/28 & 2025/10/07 \\
         \hline
    \end{tabular}
    \caption{Detailed data range for O4 run.}
    \label{tab:o4_run}
\end{table}

\begin{figure}[htbp]
    \centering

    \begin{subfigure}[b]{0.4\textwidth}
        \includegraphics[width=\textwidth]{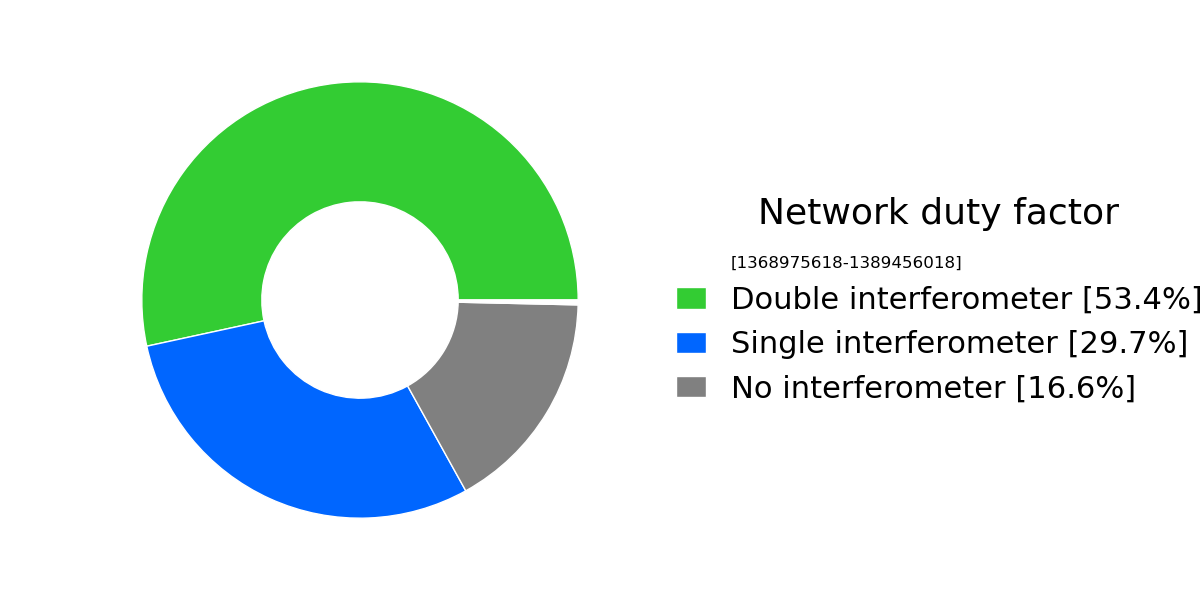}
    \end{subfigure}   
    \hfill
    \begin{subfigure}[b]{0.5\textwidth}
        \centering
        \includegraphics[width=\textwidth]{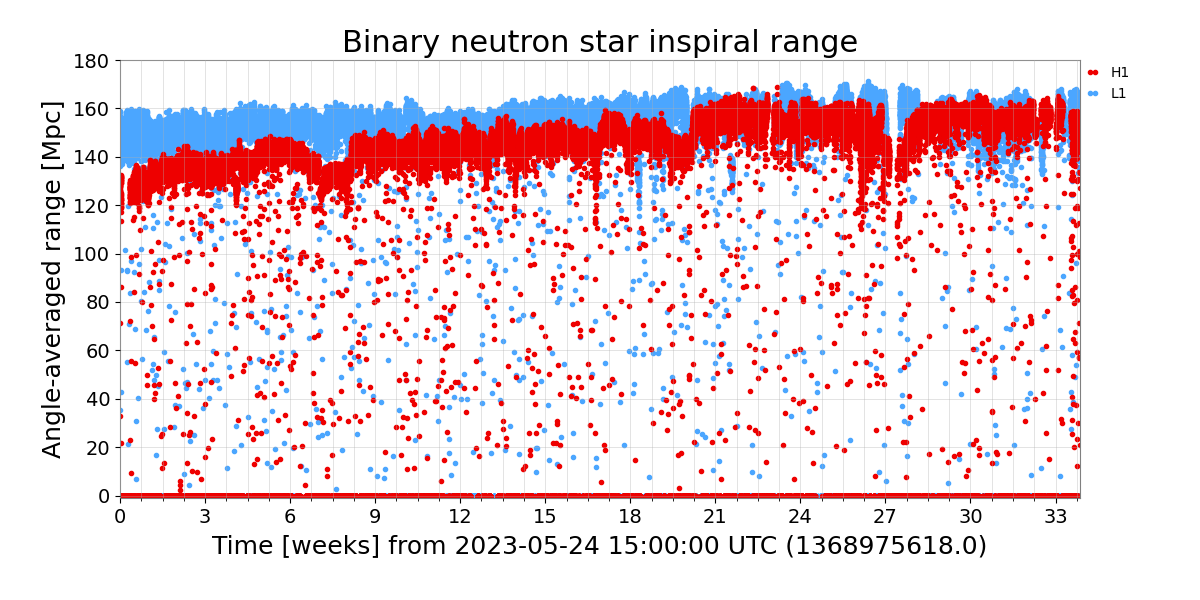}
    \end{subfigure}

    \begin{subfigure}[b]{0.4\textwidth}
        \includegraphics[width=\textwidth]{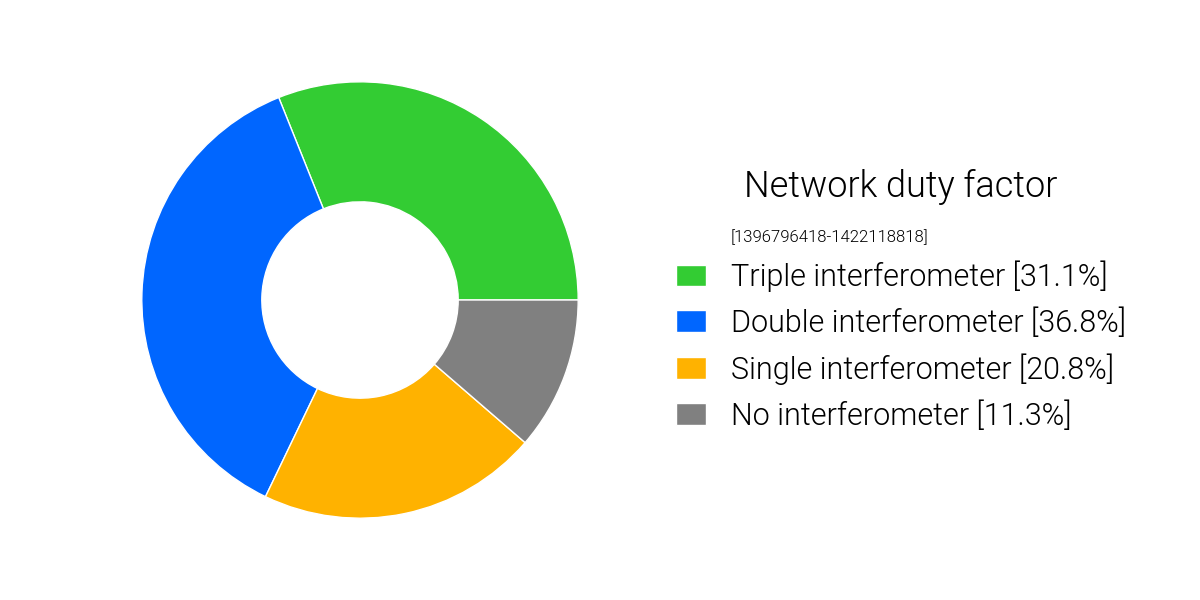}
    \end{subfigure}
    \hfill
    \begin{subfigure}[b]{0.5\textwidth}
        \includegraphics[width=\textwidth]{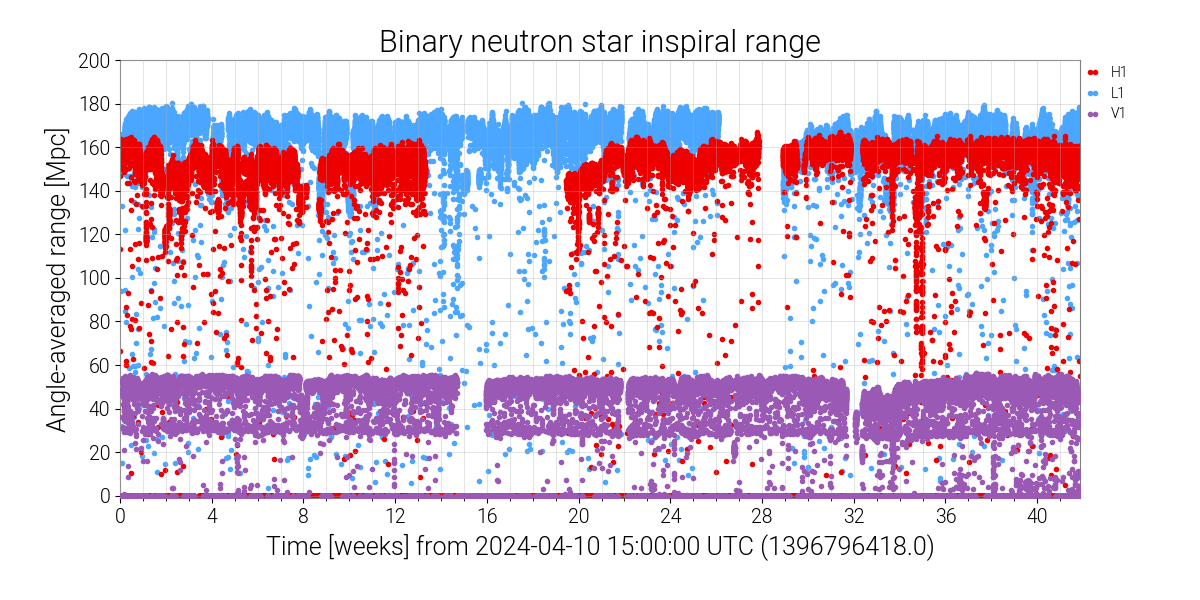}
    \end{subfigure}
 
    \begin{subfigure}[b]{0.4\textwidth}
        \includegraphics[width=\textwidth]{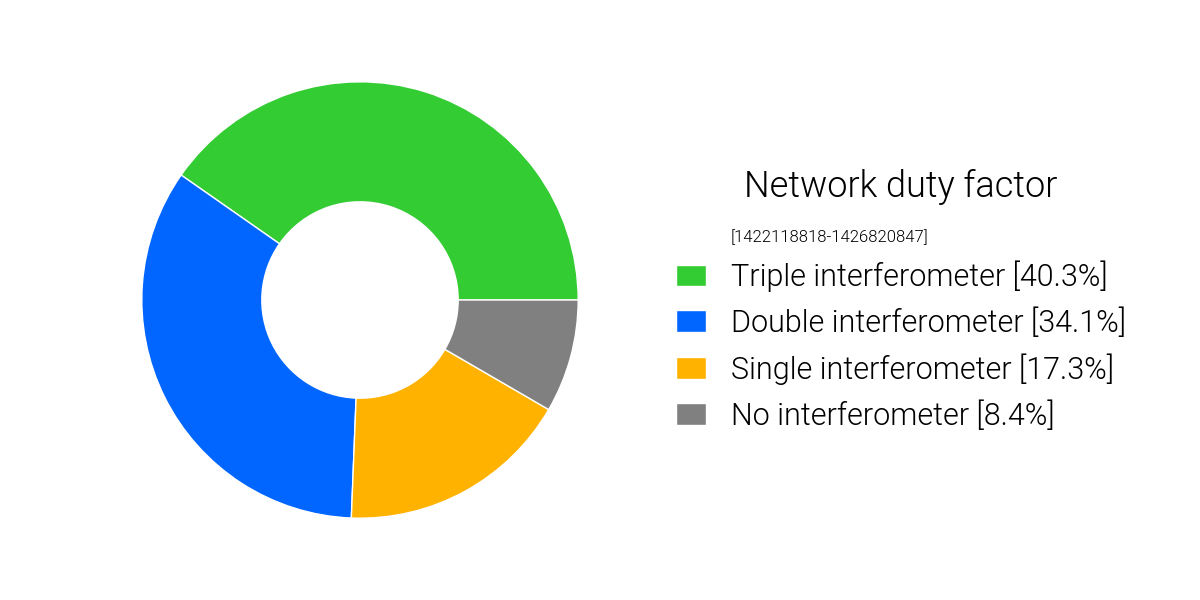}
    \end{subfigure}
    \hfill
    \begin{subfigure}[b]{0.5\textwidth}
        \centering
        \includegraphics[width=\textwidth]{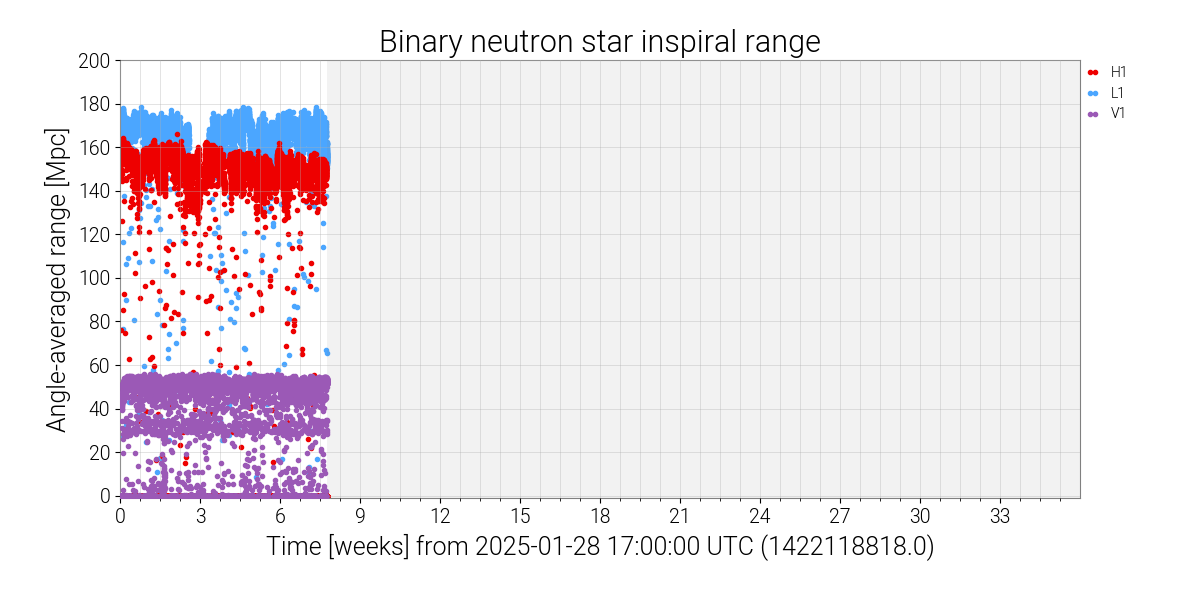}
    \end{subfigure}

    \caption{Duty factor (left) and BNS range (right) of O4a, O4b, and O4c (first, second, and third row) \protect\cite{GWOSC}.}
    \label{fig:network_o4abc}
\end{figure}

Diving into further details, there is other relevant information that helps understand the quality of a run: duty cycle and the Binary Neutron Star (BNS) range. In Table \ref{tab:virgo}, the performances for each detector are shown, while Fig. \ref{fig:network_o4abc} is a summary of the duty cycle and the BNS of the network.

\begin{table}[ht]
    \centering
    \begin{tabular}{|c|c|c|c|c|}
    \hline
    \multicolumn{5}{|c|}{\textbf{Hanford}} \\ \hline
    Run & Observing (\%) & Ready (\%) & Locked (\%) & Not locked (\%) \\ \hline
    O4a & 67.5 & 0.7 & 4.8 & 26.9 \\ \hline
    O4b & 48.6 & 0.6 & 3.5 & 47.3 \\ \hline
    O4c & 60.1 & 0.4 & 4.4 & 35.0 \\ \hline
    \end{tabular}
    \label{tab:hanford}
\end{table}

\begin{table}[htbp]
    \centering
    \begin{tabular}{|c|c|c|c|c|}
    \hline
    \multicolumn{5}{|c|}{\textbf{Livingston}} \\ \hline
    Run & Observing (\%) & Ready (\%) & Locked (\%) & Not locked (\%) \\ \hline
    O4a & 69.0 & 0.4 & 3.1 & 26.8 \\ \hline
    O4b & 68.1 & 0.6 & 5.1 & 26.0 \\ \hline
    O4c & 68.0 & 0.7 & 2.3 & 28.8 \\ \hline
    \end{tabular}
    \label{tab:livingston}
\end{table}

\begin{table}[htbp]
    \centering
    \begin{tabular}{|c|c|c|c|c|}
    \hline
    \multicolumn{5}{|c|}{\textbf{Virgo}} \\ \hline
    Run & Observing (\%) & Ready (\%) & Locked (\%) & Not locked (\%) \\ \hline
    O4b & 70.8 & --& 4.9 & 24.3 \\ \hline
    O4c & 78.2 & -- & 3.7 & 18.1 \\ \hline
    \end{tabular}
    \caption{Hanford, Livingston, and Virgo detectors operational states during O4.}
    \label{tab:virgo}
\end{table}

\newpage

\subsection{Online searches and alerts}
During the observing run, there are online pipelines that get triggered, resulting in \emph{public alerts} collected in \href{https://gracedb.ligo.org}{GraceDB}. So far, there have been 81 (O4a), 105 (O4b), and 17 (O4c) for a total of 203 alerts, surviving the retraction after further analysis. The online pipelines are divided into:

\begin{itemize}
    \item \emph{Modeled} for Compact binary coalescence (CBC): GstLAL, MBTA, PyCBC Live, and SPIIR;
    \item \emph{Unmodeled} for a wide variety of sources like core-collapse supernovae, unknown GW sources, and so on: cWB, oLIB, and MLy;
    \item For coincidences between GW and non-GW events, multi-messenger astronomy: RAVEN.
\end{itemize}

\subsection{Noise management}
The non-stationary and non-Gaussian nature of the data and the presence of noise artifacts may impact data quality or detector performance, increasing the false alarm rate of searches. Taking into account this aspect often means applying \emph{vetoes} on the acquired data, which can be broadly divided into three categories \cite{galaxies10010012}: data completely unusable and unreliable (manually removed); target small time periods around specific types of glitches (correlation with Hveto, then veto times from Omicron triggers);  based on statistical correlations (auxiliary channels vs GW strain data). According to the source of interest, different techniques are applied to the data to pursue the analysis. Moreover, in the last few years, Machine Learning approaches \cite{Cuoco_2021} have been starting to be used to further improve the data quality: GravitySpy (Deep CNN) for glitch classification; iDQ (RandomForest) from auxiliary detector channels; noise subtraction (denoisers) and many others.

\section{O4 non-CBC results}
\begin{figure}[htbp]
    \centering
    \includegraphics[width=0.45\linewidth]{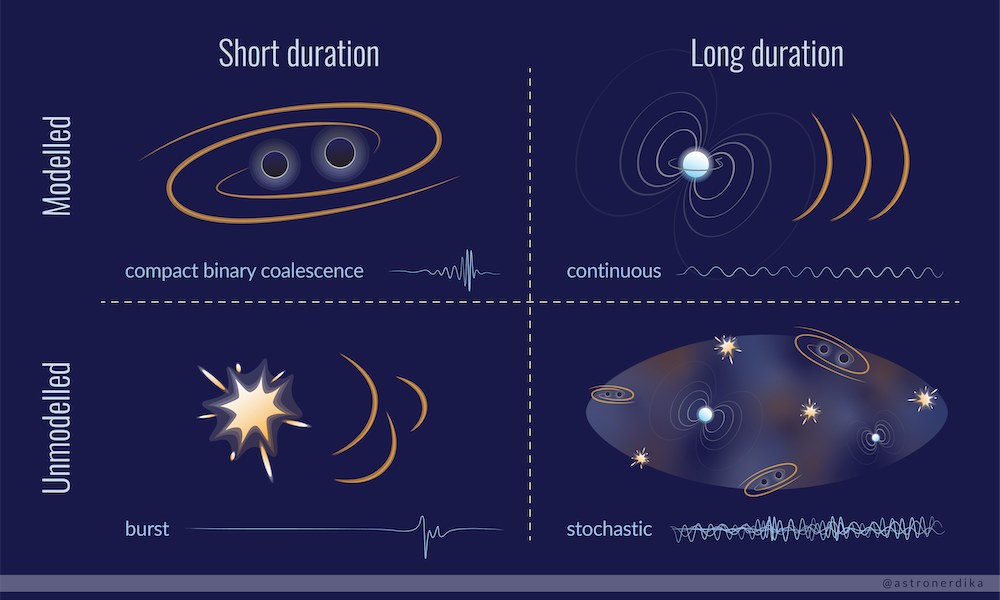}
    \caption{A broad division of the possible GW sources. Credits: Shanika Galaudage. }
    \label{fig:gw_sources}
\end{figure}

In Fig. \ref{fig:gw_sources}, a nice artistic representation of all the possible GW sources that have been looked for in the data acquired by the detectors. By crossing the horizontal and vertical axes, four broad combinations arise: CBC, Continuous, Burst, and Stochastic. The focus of the proceeding is to discuss \emph{non}-CBC results; therefore, the CBC category will be dropped off leaving room for the remaining ones.

\subsection{Continuous waves}

\subsubsection{Targeted}
\begin{figure}[ht]
    \centering
\begin{subfigure}[b]{0.45\textwidth}
        \includegraphics[width=\textwidth]{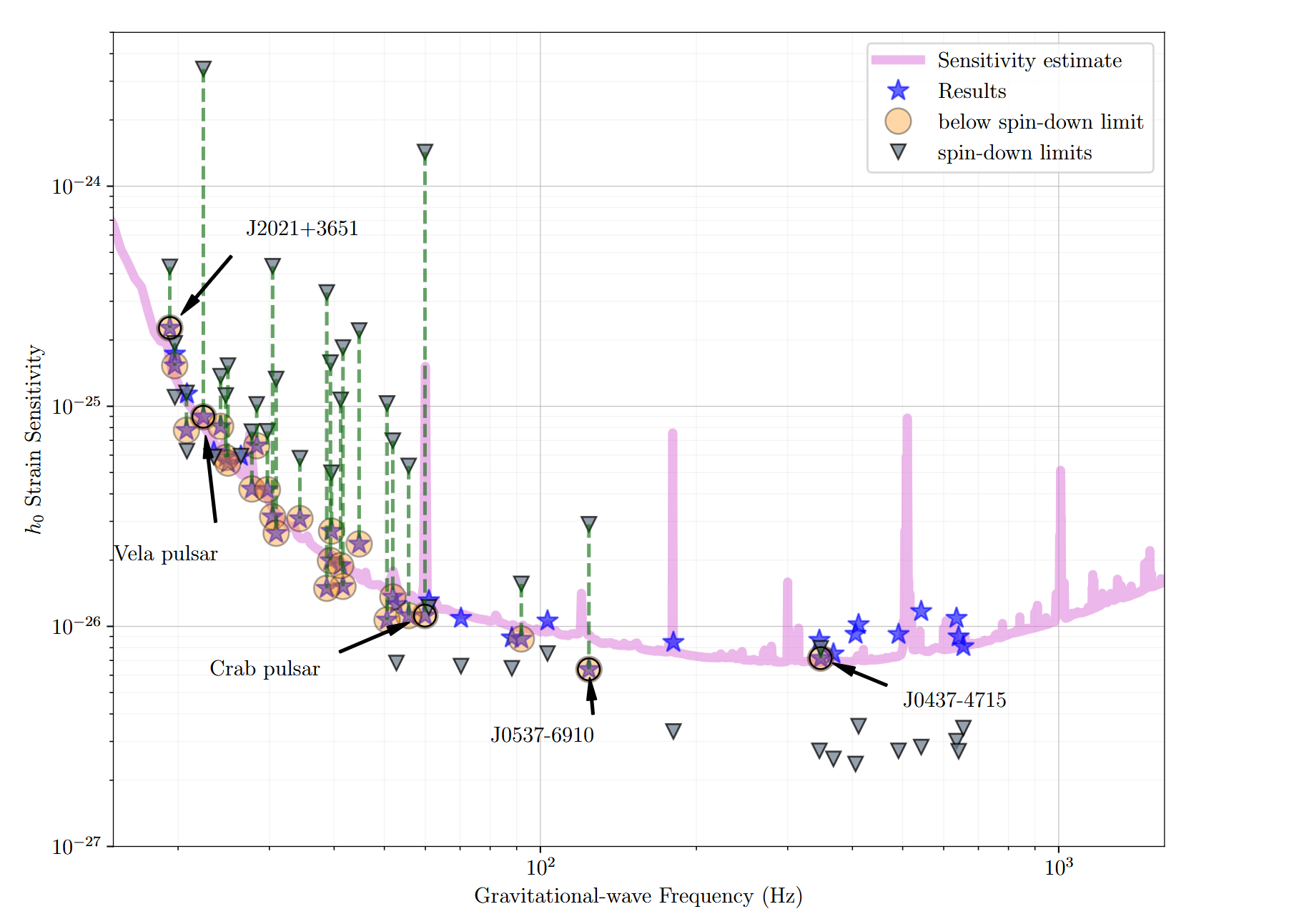}
    \end{subfigure}
    \hfill
    \begin{subfigure}[b]{0.45\textwidth}
        \includegraphics[width=\textwidth]{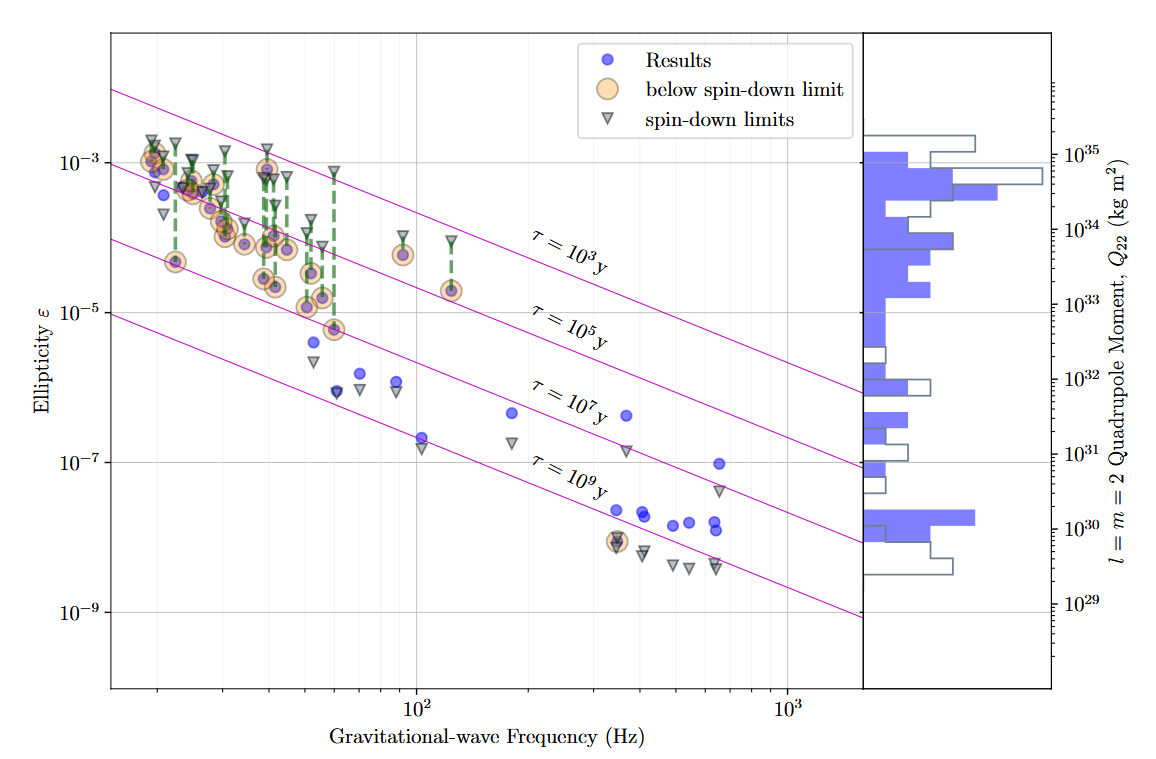}
    \end{subfigure}
    \caption{O4a Targeted search. Left panel: constraints on the strain $h$. Right panel: constraints on the ellipticity $\epsilon$.}
    \label{fig:targeted}
\end{figure}

Up to now, the only paper regarding the O4 run concerns a data analysis performed on pulsars,  specifically a \emph{targeted} search (where rotational and sky-position parameters are known) \cite{Abac_2025}. In this analysis, three different techniques are employed: Bayesian, 5-n vector, and $\mathcal{F/G}$ statistics. The plots shown in Fig. \ref{fig:targeted} concern results of the analysis run with the Bayesian approach for the strain $h_{0}$ and the ellipticity $\epsilon$, giving also info about the \emph{spin-down limit} (i.e. loss of energy entirely due to GW emission) which represents the maximum $h_{0}/\epsilon$ expected from that source. The 5-n vector method was also used on a subset of 16 pulsars for a narrowband analysis.\\ The authors of the paper conclude that no GW signal was found in either of the three searches.

\subsubsection{All-sky}
\begin{figure}[htbp]
    \centering
\begin{subfigure}[b]{0.45\textwidth}
        \includegraphics[width=\textwidth]{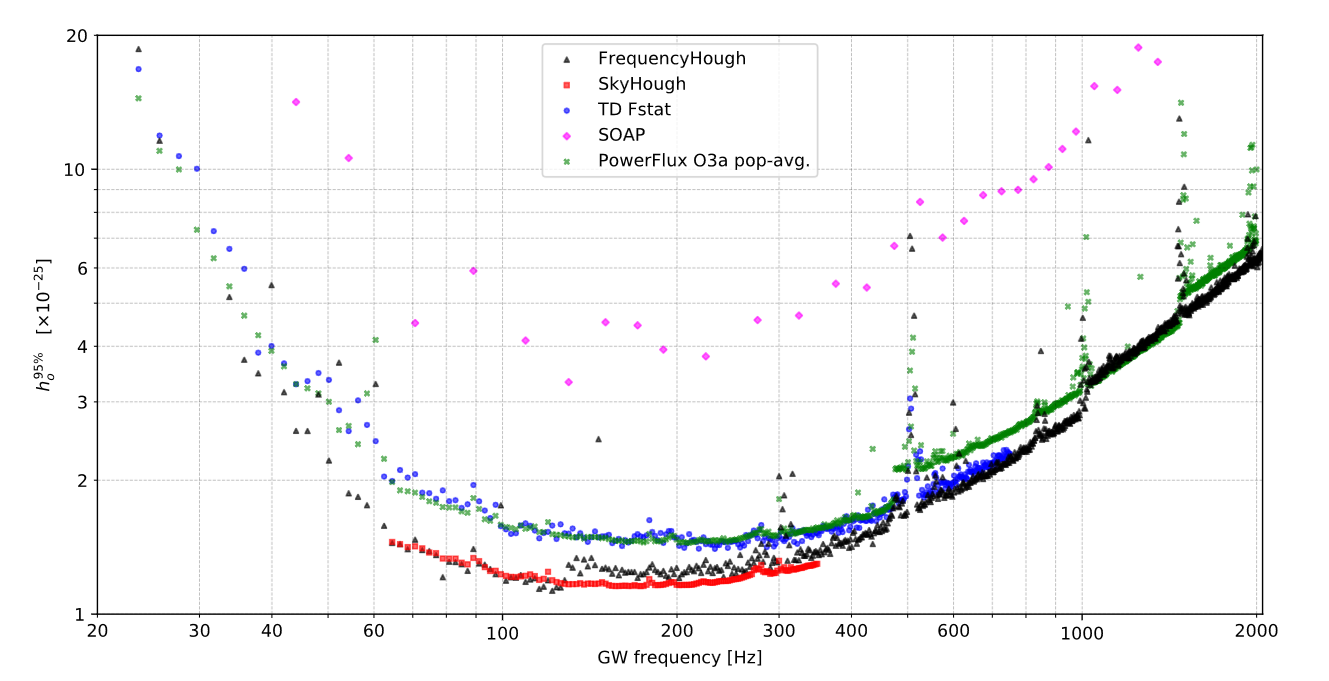}
    \end{subfigure}
    \hfill
    \begin{subfigure}[b]{0.45\textwidth}
        \includegraphics[width=\textwidth]{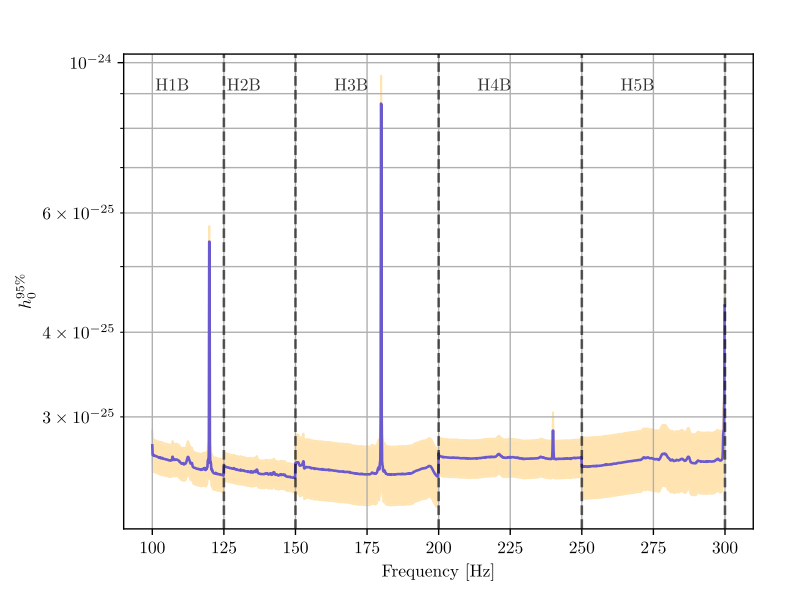}
    \end{subfigure}
    \caption{Latest O3 results for All sky searches. Left panel: isolated neutron stars. Right panel: neutron star in a binary system.}
    \label{fig:all_sky}
\end{figure}

For the All-sky search, no parameter of the source is known, making it quite computationally expensive to perform. In Fig. \ref{fig:all_sky} there are summarized different upper limits (ULs) for the isolated  \cite{PhysRevD.106.102008} and binary \cite{PhysRevD.103.064017} neutron star scenarios. For the former, different methods are applied: \texttt{FrequencyHough, SkyHough, Time-Domain $\mathcal{F}$-statistic, SOAP}. The best value is $h_{0}=1.1\cdot 10^{-25}$, reached at $f=111.5$ Hz. The latter is performed by \texttt{BinarySkyHough}, resulting in $h_{0}^{95\%}=(2.4\pm0.1)\cdot10^{-25}$ as the lowest value at $f=149.5$ Hz.

\subsubsection{New physics}
\begin{figure}[htbp]
\centering
    \includegraphics[width=0.45\textwidth]{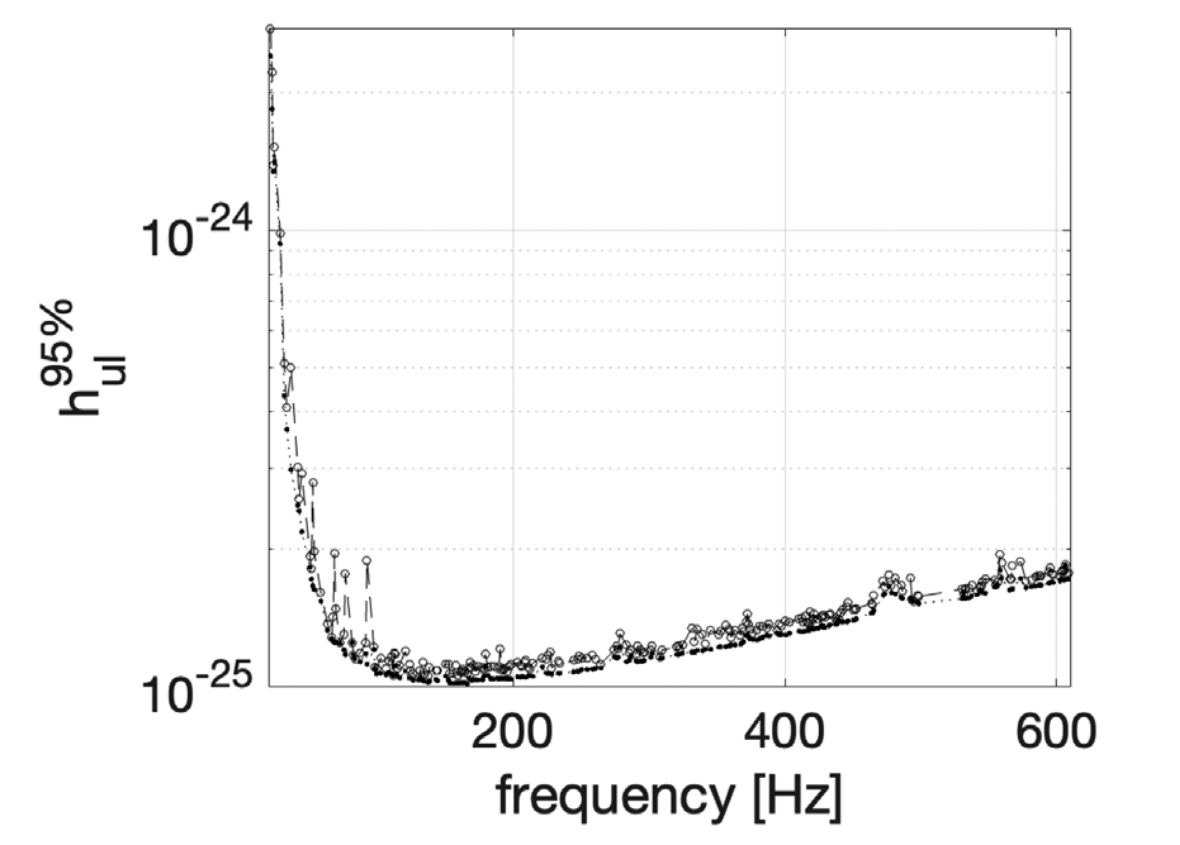}
    \caption{Upper limit for GW from boson clouds.}
    \label{fig:boson_UL}
\end{figure}

\begin{figure}[htbp]
\centering
    \includegraphics[width=0.6\textwidth]{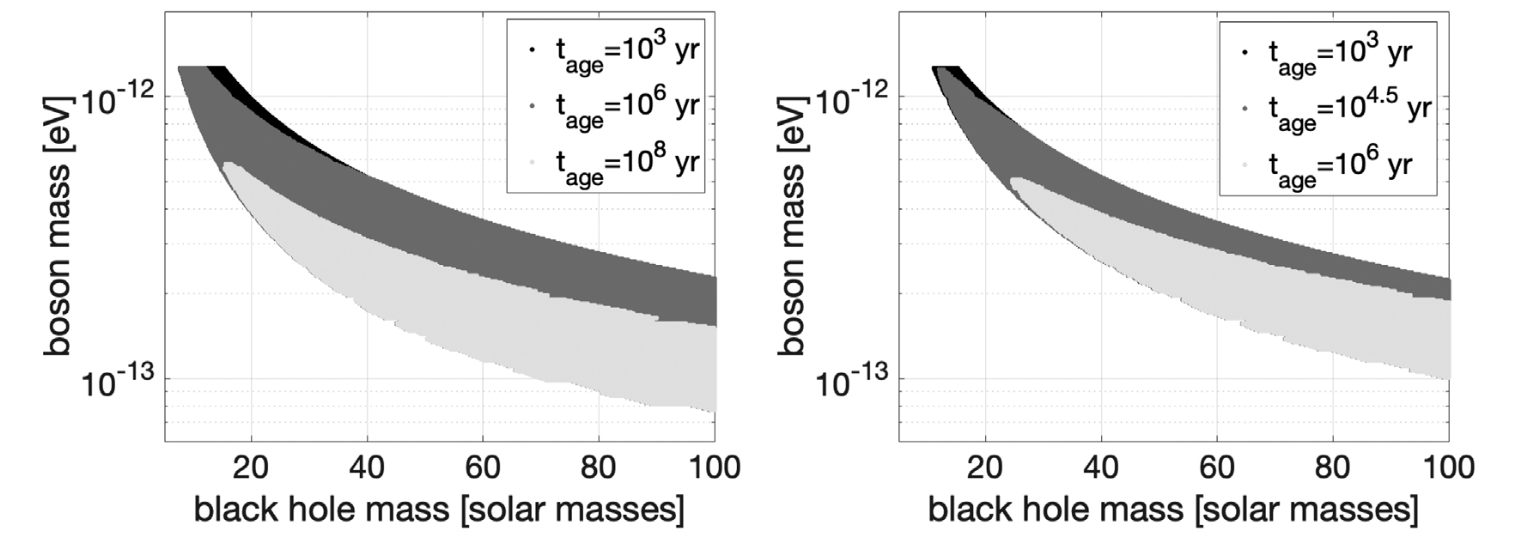}
    \caption{Exclusion regions for black hole and boson masses.}
    \label{fig:boson_ex}
\end{figure}

An interesting search for Continuous waves regards considering a boson cloud as a possible  GW source \cite{PhysRevD.105.102001}. Here, ultralight bosons (dark matter, and beyond the Standard Model) - around a black hole - would go through annihilation into gravitons, and the spin up from the contraction of the cloud as it loses mass would result in GW signal. In Fig. \ref{fig:boson_UL}, the upper limit obtained has its best value $h_{0}^{95\%}=1.04\cdot10^{-25}$ at $f=130.5$ Hz. From the UL, exclusion region of the black hole and boson masses can be extracted, as shown in Fig. \ref{fig:boson_ex}.

\subsection{Burst}
\begin{figure}[htbp]
    \centering
    \includegraphics[width=0.5\linewidth]{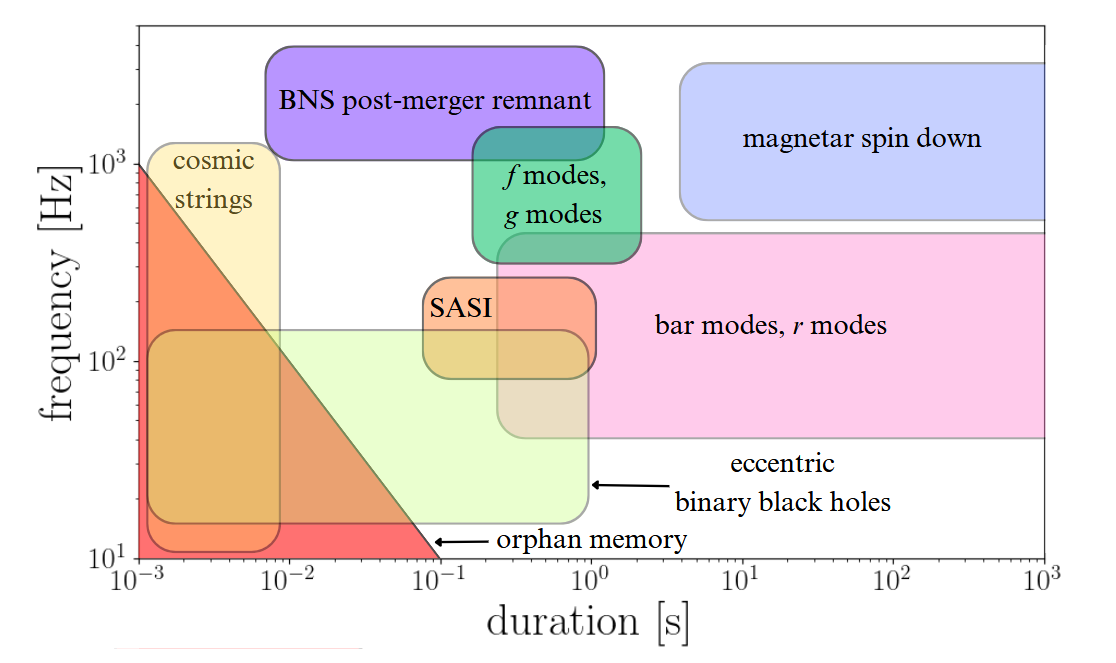}
    \caption{Different sources for Burst signals.}
    \label{fig:burst_overview}
\end{figure}

Many sources fall under the umbrella of \emph{Burst} signals, as it can be seen in Fig. \ref{fig:burst_overview}.
They can be broadly divided into \cite{powell2025dawesreviewgravitationalwaveburst}:
\begin{itemize}
    \item CCSNe: complex scenario to model due to multiple aspects to take into account (hydrodynamics, EoS, ...). The current available waveforms can distinguish
generation mechanisms with SNR$>$20, $g/f$ modes are the strongest features. $15\leq \rm{SNR}\leq20$ for unmodeled searches.
    \item BNS remnants: Stable, hypermassive, supramassive NS $\leq$ 1 s after the merger, kHz regime. Specific frequencies for EoS. More long-lived NSs emit GW for longer periods of time. Not enough BNS events to determine mass distribution. Moreover, at least one order of magnitude of additional sensitivity from detectors is needed;
    \item Eccentric binaries: Eccentric systems circularize due to GW emission. For $e \geq 0.5$ they resemble short-lived bursts just before the merger. Standard matched filter searches do not include eccentricity (but they are still effective for $e \leq 0.2$). LVK uses cWB for unmodelled burst signals;
    \item FRBs: Astrophysical progenitor often unknown. Non-repeating: BNS proposed explanation. Modeled and Unmodeled searches. Only unmodeled for repeating FRBs;
    \item Pulsar glitches: A sudden increase in the angular moment causes an increase in the spin frequency and GW. Lag between core and crust. Mechanisms: Spin-up event rings up many oscillation modes, short lived. Relaxation of the star interior, GW through the current quadrupole. $f/p$ modes for GW, short-lived ($\leq0.1$ s). Glitch models constrained by non-observations;
    \item Magnetars: NSs with $B\geq 10^{13}$ G. Crust-cracking, internal magnetic fields rearrangements could generate GW through $f$-modes oscillations (short-lived) or $g$-modes (long-lived)/Alfven waves. LVK searches for GW from magnetar flares. UL on GW energy $\leq 10^{47}$ erg. UL for $f/g$-modes + Alven: $\leq 10^{44}$ erg;
    \item GRs: Long ($\geq$ 2 s), extreme SNe explosion. Time window: [- 600, 60] s w.r.t. GRB event;
    \item Orphan memory: from anisotropic emission of GW. Non-existence of OM bursts in LVK data used to constrain models in MHz and GHz;
    \item and many others...
\end{itemize}

From all the analysis performed for the different sources of \emph{burst}, no GW signal was found.

\subsection{Stochastic}
\begin{figure}[htbp]
    \centering
\begin{subfigure}[b]{0.4\textwidth}
        \includegraphics[width=\textwidth]{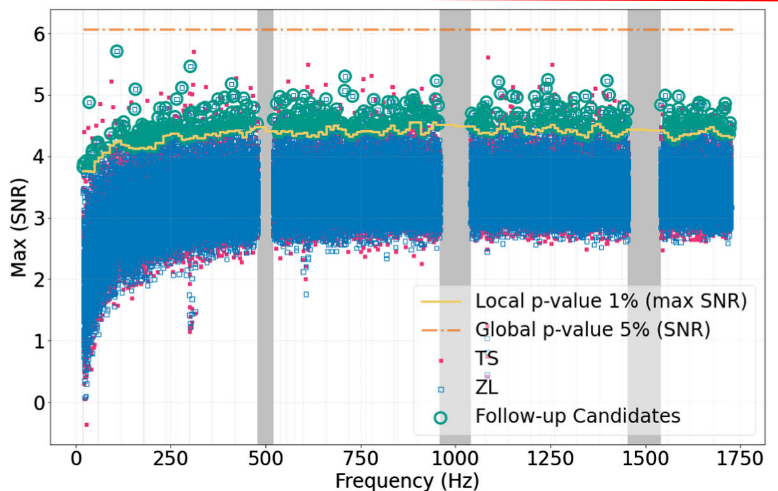}
    \end{subfigure}
    \hfill
    \begin{subfigure}[b]{0.4\textwidth}
        \includegraphics[width=\textwidth]{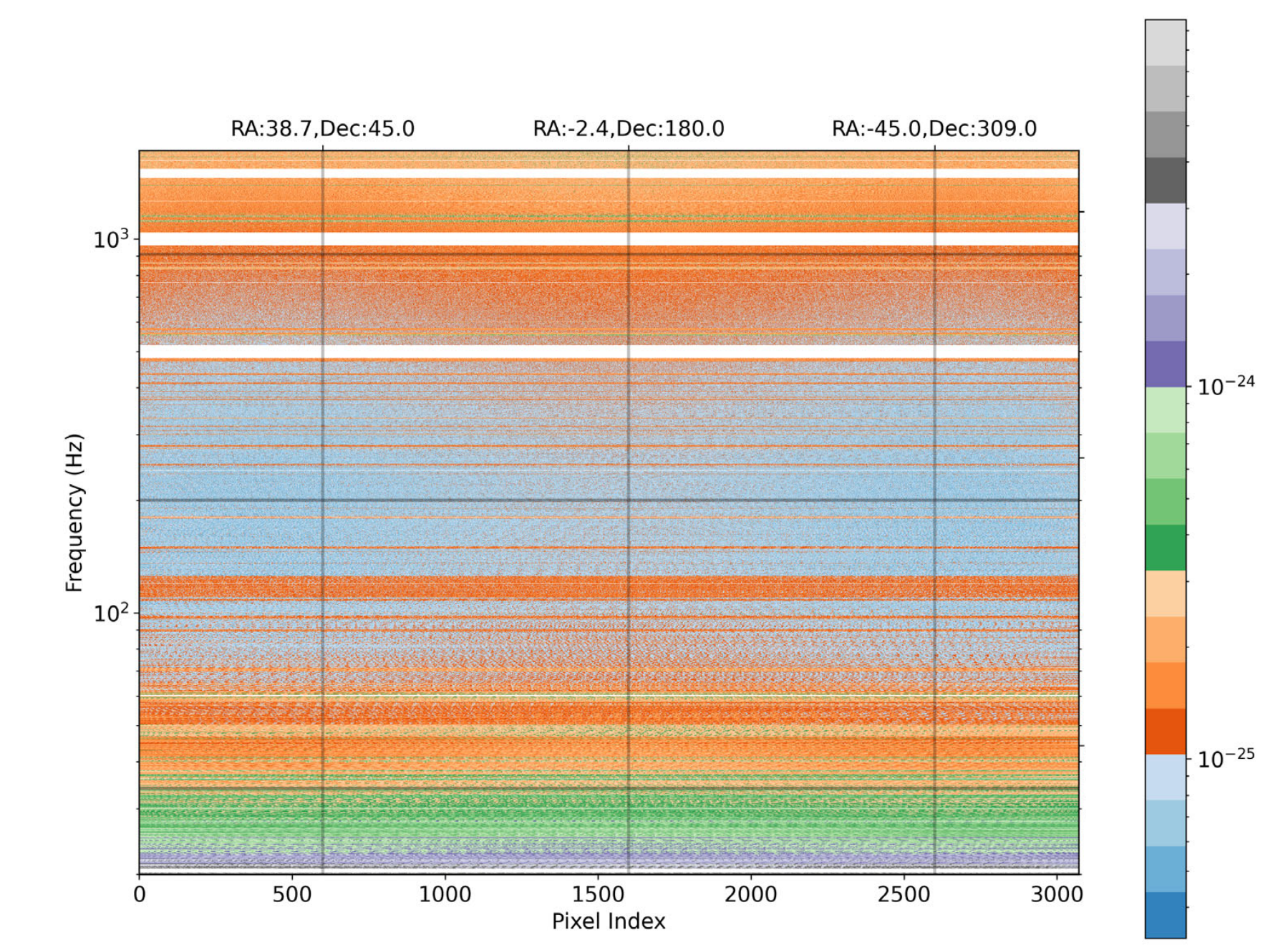}
    \end{subfigure}
    \caption{Left panel: Candidates from O1+O2+O3 data. Right panel: UL at different frequencies and sky locations.}
    \label{fig:stochastic}
\end{figure}

All-sky-all-frequency (ASAF) \cite{PhysRevD.105.122001} is an analysis that looks for the Stochastic GW background (SGWB). The procedure is first validated with O2 + Hardware injections (HIs) to be recovered; then tested with the combination of all the runs so far, O1+O2+O3, and no injections, looking again for outliers. The result of the analysis is shown in Fig. \ref{fig:stochastic}, left panel: since no Follow-up candidates are above the dashed orange line, which represents the $p$-value for an interesting candidate, no GW was found in the data. After this conclusion, the method was applied for ULs at different frequencies and pixel index (which is related to the sky location due to the use of HEALPix), right panel.

\newpage

\section{Conclusions}
The O4 has started on 24$^{th}$ May 2023 and it will end on 7$^{th}$ October 2025. Although experiencing a few issues during the acquisition period (earthquake for KAGRA, mystery noise in Virgo, and detector failure for LIGO), the overall sensitivity of the network is the best so far, reaching the huge milestone of $>$200 public alerts thanks to the multiple online pipelines working in real time. Moreover, different noise management techniques are applied to improve the data quality according to the branch of interest, also implementing ML methodologies.\\
The entire community analyzes data to look for GW signals from the Universe. Concerning the O4 run, only one non-CBC paper has been published: for continuous signals, targeted strategies have been applied for known pulsars, getting new limits on the strain $h_{0}$ and the ellipticity $\epsilon$. Apart from it, the latest O3 results have been presented: Continuous, Burst, and Stochastic searches have been able to obtain UL since no GW signal of any of this nature has been found in the data. More investigations are ongoing.

\section*{References}
\bibliography{references}

\end{document}